\documentclass[nolinenumbers]{ametsocV6.1}

\usepackage{amssymb}
\usepackage{appendix}
\usepackage{natbib}
\usepackage{graphicx} 
\nolinenumbers

\bibpunct{(}{)}{;}{a}{}{,}

\title{Explainable Deep Learning for Probabilistic Nowcasting of Radar Reflectivity in Tornadic Storms}
\authors{Nathan Erickson\aff{1,2}\correspondingauthor{Nathan Erickson, nathan.erickson@ou.edu},
Amy McGovern\aff{1,2,3}, Aaron Hill\aff{1,2}}

\affiliation{\aff{1}{University of Oklahoma, School of Meteorology, Norman, OK}\\
\aff{2}{National Science Foundation Institute for Research on Trustworthy AI in Weather, Climate, and Coastal Oceanography, Norman, OK}\\
\aff{3}{University of Oklahoma, School of Computer Science, Norman, OK}\\}

\begin{document}

\nolinenumbers
\abstract{Tornadoes pose substantial risk to human life and property in the United States, causing more than 50 fatalities and \$100 million of property damage on average annually. When tornadoes are likely, weather radar provides critical information for forecasters by providing information on storm morphology, storm motion, and intensity
trends. Additional tools such as satellite and numerical weather prediction model runs can provide useful short-term information for understanding changes in storm characteristics. This work demonstrates a U-Net deep-learning system for nowcasting the evolution of radar reflectivity following tornadogenesis, which can provide value to forecasters
by synthesizing large amounts of input data (e.g., radar imagery, near-storm environment data) and generating predictions of radar reflectivity from its inputs. Inputs to the model are radar imagery from the Multi-Radar Multi-Sensor (MRMS) dataset and near-storm environment data from the High-Resolution Rapid Refresh (HRRR) numerical weather prediction model. The U-Net is trained on a dataset of tornadic storms to produce 30 minutes of probabilistic predictions of radar reflectivity following tornadogenesis, with probabilistic predictions obtained by predicting parameters of the SinhArcSinh, or SHASH, distribution. The model produces physically realistic predictions of radar evolution (achieving a domain-wide RMSE of 7.71 dBZ), achieves comparable skill to next-hour forecasts from the HRRR (0.73 FSS on a 16x16 km window for the U-Net vs. 0.69 for the HRRR for reflectivity forecasts 15 minutes following tornadogenesis), demonstrates reasonable probabilistic calibration (with the model achieving a PITD-statistic of 0.3) and is accompanied by a variety of explainability methods to improve understanding by end users. Additionally, predictions from the model can be obtained much more quickly than those from a numerical weather prediction model (30-minute prediction for one tornadic storm can be achieved in approximately 1 second on an NVIDIA A100 GPU). Drawbacks to the model include blurring predictions with lead time and a reduced representation of smaller-scale storm structures. With further development, this model could be extended to nowcast radar reflectivity evolution in an operational setting.}

\maketitle

\statement
Radar is a critical tool forecasters rely on when issuing tornado warnings. However, predicting radar evolution is very challenging. This paper introduces an AI model to produce predictions of radar evolution in the short term (i.e., out to 30 minutes). Our deep learning model produces forecasts on par with those from current state-of-the-art weather models while using far fewer computational resources. With further improvements, this model can support forecasters when issuing tornado warnings by providing accurate estimates of radar evolution.

\section{Introduction}
Understanding and predicting the behavior of tornadic storms is critical for the protection of life and property \citep{NCEITornadoes}. 
Recently, the increasing availability of data from meteorological instruments (e.g., radar and satellite), as well as from numerical weather prediction (NWP) models (i.e., near-storm environment data) has improved the prediction of impacts from severe convective storms. Among these products, radar has emerged as critical to the operational workflow for forecasting tornadoes, being used to issue timely warning products and interpret storm characteristics such as intensity, morphology, and motion in near-real-time \citep{NOAARadar, WeberRadar}. However, in spite of its usefulness in operations, radar is not a prognostic tool - while it can indicate possible future storm behavior through development of a hook echo or a rear-flank downdraft surge, it cannot directly predict future storm characteristics. 



In an attempt to extend radar capabilities, a variety of approaches have been employed to enable radar nowcasting, or the production of very short-term forecasts; see \citep{MuellerAutoNowcast, RobertsAndRutledgeNowcasting, MecikalskiConvectiveInitiation, SteinheimerAndHaiden, PruddenRadarNowcasting} for an in-depth discussion of radar nowcasting techniques. Perhaps the simplest approach is to employ persistence-based nowcasting \citep{GermannAndZawadzkiPersistenceNowcasting}. Persistence can be Eulerian (using the present state as the prediction for future states), Lagrangian (assuming that the only change to the field is due to advection), or based on the persistence of individual convective cells and their characteristics. Probabilistic and stochastic approaches such as PySTEPS expand on the persistence-based method; probabilistic methods do so by modeling a probability distribution of the reflectivity field to account for intrinsic error, while stochastic approaches inject spatially-correlated noise to model intrinsic error \citep{SeedNowcasting2003, BowlerSTEPS, SeedNowcasting2013}. However, these approaches still may fail to capture the evolution of the advection field, or changes in the underlying convective regime itself.

Existing baselines often build on persistence-approaches to extrapolate the current radar state. One such approach for persistence-style nowcasting is the optical flow method \citep{BechiniOpticalFlow}, which estimates the motion of an object from sequentially-ordered images. \citet{BechiniOpticalFlow} established an average threat score of 0.3 at a 30-minute lead time when forecasting precipitation using an optical flow technique, indicating some skill from this method. \citet{ImhoffExtrapolationNowcasting} evaluated PySTEPs and Rainymotion benchmark algorithms relative to existing Eulerian persistence baselines, finding superior performance from the PySTEPS algorithm compared to the other benchmark approaches. Still, while persistence-based approaches may be useful for forecasting the evolution of large-scale precipitation fields, it is unlikely to provide sufficiently high skill for accurately predicting changes in rapidly evolving convective storms.

NWP models are also employed to generate severe weather forecasts at nowcasting timescales. Perhaps the most widely-used tool for predicting convective development at nowcasting timescales is the Warn-On Forecast System [WoFS;  \cite{StensrudWoFS}), an NWP ensemble of 36 members (18 forecast members] which utilizes 15-minute assimilation of observations, including real-time radar and satellite data, to generate rapidly updating predictions at convective scales. WoFS generates skillful forecasts for many types of convective weather, including severe convective storms \citep{SkinnerWoFS,ClarkAndLokenWoFS,HeinselmanWoFS2024}, tropical cyclone hazards \citep{JonesWoFSTC, YussoufWoFSTC}, and flash flooding \citep{YussoufAndKnopfmeier}. However, WoFS still suffers from some of the drawbacks of traditional NWP models, for example, requiring 30 wall-clock minutes to generate a 6-hour forecast \citep{HeinselmanWoFS2024}. When nowcasting radar, rapid forecast production is a desirable quality.

An alternative to traditional NWP models is deep learning. Deep learning, a subset of machine learning, utilizes neural networks to algorithmically learn complex, non-linear mathematical functions from a training dataset. Deep learning models have been used to study topics in convective meteorology such as convective initiation/development \citep{LagerquistCI,LagerquistAndEbertUphoff}, hail \citep{WangHail, PullmanHail, SpychallaHail, SchmidtHail}, straight-line wind \citep{MounierWind, CoburnWind}, and tornado detection/prediction \citep{MarzbanTornado, LakshmananTornado, LagerquistTornado}. Deep learning provides multiple advantages compared to traditional NWP models, such as automatic discovery of non-linear functions that are not explicitly encoded in the model and fast inference time (see \citet{ChaseUVV} for a discussion of inference time vs. training time). 

Previous studies involving nowcasting with deep learning have achieved comparable skill to baseline NWP and statistical techniques. WoFSCast was developed as a machine-learning-based emulator of WoFS; it generates deterministic forecasts of atmospheric state variables out to 2 hours at convection-allowing scales while achieving comparable performance to WoFS \citep{WoFSCast}. WoFSCast provides dramatic improvement in the time and computational resources required to generate a forecast when compared to the physics-based WoFS ensemble; WoFSCast predictions can be generated in 30-40 seconds on a single GPU, as compared to the thousands of CPU cores required to generate WoFS ensemble forecasts over the course of 30 minutes. Similarly, HRRRCast \citep{HRRRCast} was trained as an emulator of the NOAA High-Resolution Rapid Refresh [HRRR; \citet{DowellHRRR}]; HRRRCast achieves comparable performance with the HRRR by training a diffusion-based model on HRRR analysis data. NVIDIA's StormScope \citep{StormScope} is another example of a competitively-performing high-resolution NWP model emulator, offering forecasts at 10 minute temporal frequency and 6 km horizontal grid spacing at lead times of up to 6 hours.

For nowcasting radar specifically, \citet{ChenDLNowcasting} used a convolutional long short-term memory (LSTM) model to generate skillful predictions of radar reflectivity at 30 and 60 minute lead times. \citet{RavuriDLNowcasting} employed a deep generative model to produce precipitation predictions for heavy rain events. \citet{CuomoDLNowcasting} used a variety of convolutional and recurrent model architectures and achieved comparable skill to baseline (i.e., non-deep learning) statistical models.  However, a component notably lacking from many studies on deep learning for radar nowcasting is any form of uncertainty quantification (UQ); \citet{RavuriDLNowcasting} is a notable exception in this regard. UQ enables the consideration of multiple potential outcomes from a prediction, rather than only a single deterministic outcome. Probabilistic information from UQ is critical in the atmospheric sciences to guide decision making and evaluate risk \citep{MurphyAndKatzUQ, BarnesUQ}. 

We employ a deep learning model known as a U-Net \citep{RonnebergerUNet} to generate predictions of radar evolution in tornadic storms out to 30 minutes of lead time. While models such as WoFSCast and StormScope produce general forecasts of the atmospheric state at convective time scales, we produce a radar-nowcasting model intended to generate very-short-term forecasts of radar reflectivity specifically. We use a novel method for the prediction of radar reflectivity evolution in a probabilistic framework: the U-Net is trained to predict the parameters of the SinhArcsinh (SHASH) probability distribution \citep[e.g.,][]{BarnesSHASH, ChaseUVV}. We hypothesize that a U-Net trained on prior radar imagery and near-storm environment data from the previous hour will be able to produce skillful probabilistic predictions of radar reflectivity relative to the HRRR and existing nowcasting baselines by leveraging physical information from radar and NWP datasets, while associated explainability techniques can provide understanding on the relevant regions and variables for informing the model's predictions.

\section{Data}

\subsection{MRMS}
To create a model capable of nowcasting radar reflectivity evolution, ground truth radar reflectivity is needed to serve as training inputs and targets. For this purpose, we use the Multi-Radar Multi-Sensor dataset, version 12 [MRMS; \citet{LakshmananMRMSSvr, ZhangMRMSQPE}]. MRMS is a high-resolution gridded dataset containing over 100 radar products across the U.S. We utilize the composite reflectivity mosaic (full name \textit{MergedReflectivityQCComposite\_00.50}, or MRQC) product as reflectivity inputs for the deep learning model. MRQC has standard reflectivity units of Z (decibels relative to Z, the equivalent reflectivity factor). Missing and out-of-range values are indicated with -99 and -999, respectively, and these are masked prior to constructing the model-ready dataset. We use the MRQC product to mitigate the impacts of beam blockage in regions with terrain near the radar and to make better use of radar data both very close to the radar (where low-level scans may not intersect with high-based storms) and very far from the radar (where resolution is naturally lower, thus lowering the amount of data available at a given tilt). While using the composite reflectivity may limit operational capabilities of a model trained on such data, we believe the trade-offs are worthwhile, and similar models could be trained on base reflectivity in the future. MRQC is available beginning in October 2020 at 2-minute temporal frequency and 1-kilometer grid spacing across CONUS. MRQC serves as both input data and target data for the deep learning model (further information on case construction can be found in Section 3a).

\subsection{HRRR}
In addition to radar inputs from MRMS, we also incorporate near-storm environment data from the HRRR model to supply additional relevant features for radar evolution. The HRRR model is an operational convection-allowing numerical weather prediction model, initialized once per hour at 3-kilometer horizontal grid spacing to generate high-resolution forecasts across CONUS. The HRRR assimilates radar data every 15 minutes, which can improve predictions of atmospheric phenomena such as severe convective storms, smoke plumes, etc. The HRRR provides forecasts of atmospheric variables on pressure levels and at specific levels/layers of the atmosphere. Namely, it offers 14 variables on 51 pressure levels, with an additional 77 surface/height level/atmospheric variables (some of which are computed directly through the governing equations, and some of which are derived). To generate predictors, we use variables from the previous-hour HRRR analysis rather than the forecast fields. While using analysis from the previous hour may result in some offsets between storm location in the analysis and at the time of tornadogenesis, the HRRR analysis still provides the best approximation of the near-storm environment in real time and would be most representative of the near-storm-environment data available to forecasters in an operational setting.

\subsection{Storm Events}
Tornadic cases are selected from the National Centers for Environmental Information (NCEI) Storm Events dataset \citep{NCEIStormEvents}. This database provides a wide array of information on all documented severe weather events from January 1950 through the present. The archive includes an accurate and quality controlled record of the start time/location of tornadoes, which provides a starting point for constructing cases centered around tornadic storms. 


\section{Methods}

\subsection{Dataset Construction}
From Storm Events, we have a baseline dataset of nearly 5,000 tornadic cases from October 2020 through December 2024. Once cases have been identified, inputs and prediction targets are selected on a 128x128 grid point box; with MRMS' 1 km grid spacing, this translates to 128 km on a side. Inputs to the deep learning model consist of 15 time steps (i.e., 30 minutes) of MRQC imagery prior to tornadogenesis and 10 HRRR variables (detailed below in Table 1), resulting in a total of 25 predictors for the model. The chosen HRRR variables are selected as characteristics of the thermodynamic, kinematic, and near-surface fields surrounding the tornadic storm. These variables are regridded from their native 3 km grid to the 1 km grid of MRMS using bilinear interpolation. All inputs to the model are normalized using min-max normalization.

\begin{figure}
    \centering
    \includegraphics[width=0.7\linewidth]{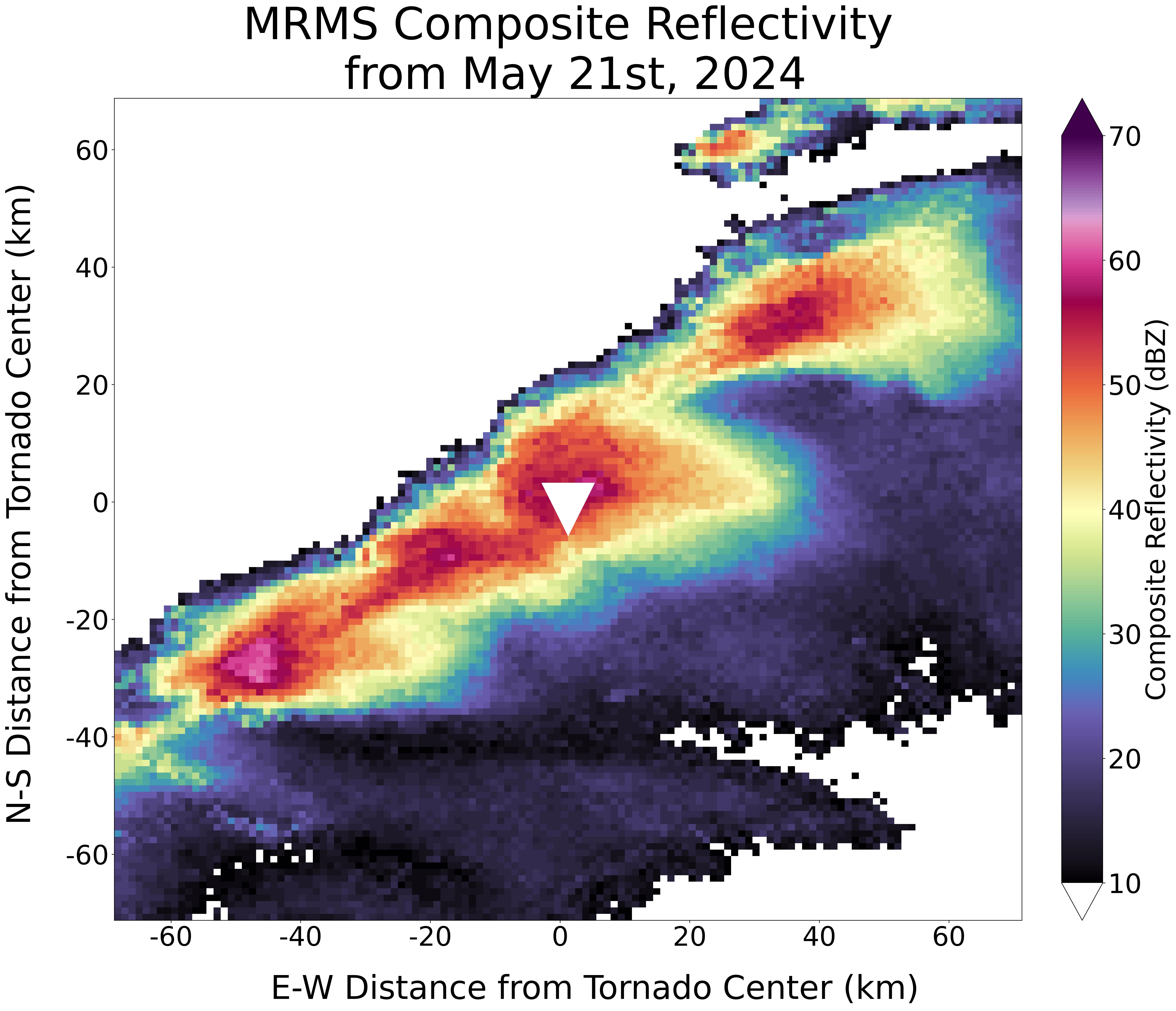}
    \caption{Example radar reflectivity input for model training, shown from training case on May 21st, 2024. Starting location of tornado is marked in white.}
    \label{fig:figure1}
\end{figure}

The target being predicted by the model consists of 15 time steps of MRQC following tornadogenesis, also on a 128x128 grid box; an example target image for one time step is shown in Figure~\ref{fig:figure1}.


\begin{table}
    \centering
    \begin{tabular}{c|c}
         HRRR Predictor&Atmospheric Level/Layer \\
         \hline
         U-component of wind shear ($m s^{-1}$)& 0-1 km\\
         V-component of wind shear ($m s^{-1}$)& 0-1 km\\
         U-component of storm motion ($m s^{-1}$)& 0-6 km\\
         V-component of storm motion ($m s^{-1}$)& 0-6 km\\
         Storm-relative helicity ($m^2 s^{-2}$)& 0-1 km\\
         Air temperature ($K$)& 925 mb\\
         Dew point temperature ($K$)& 925 mb\\
         U-component of wind($m s^{-1}$)& 500 mb\\
         V-component of wind($m s^{-1}$)& 500 mb\\
         Mixed-layer CAPE ($J  kg^{-1}$) & Lowest 100 mb\\
    \end{tabular}
    \caption{List of HRRR predictors used as model inputs, along with the atmospheric level/layer from which they are taken.}
    \label{tab:table1}
\end{table}

To increase the data volume available for model training, we perform data augmentation by applying random translations around the tornadic event to generate four unique samples for each case. The first unique sample is always created with the center of the images on the starting location of the tornado. Subsequent samples are generated by applying a consistent random shift in latitude and longitude to all input and target images. The latitude/longitude shifts are randomly generated independently of one another, and can range from 0 to 96 km in each direction to ensure that the tornadic storm remains within the image even with the latitude/longitude shifts. By augmenting our dataset, we provide a more diverse set of training data for the model to learn from, enabling it to derive more robust conclusions about the reflectivity fields. After performing data augmentation, we have a training size of 18,389 unique input/target sets. Of these unique cases, we use 2/3 of the dataset for training, 1/6 for validation, and an additional 1/6 for testing. The training/validation/testing splits are computed by sorting all of the input/target sets by date, and then splitting the dataset accordingly. Sorting and then splitting the dataset in this manner avoids cross-contamination at inference time (i.e. no storms in the test set are seen by the model during training).

\begin{figure}
    \centering
    \includegraphics[width=\linewidth]{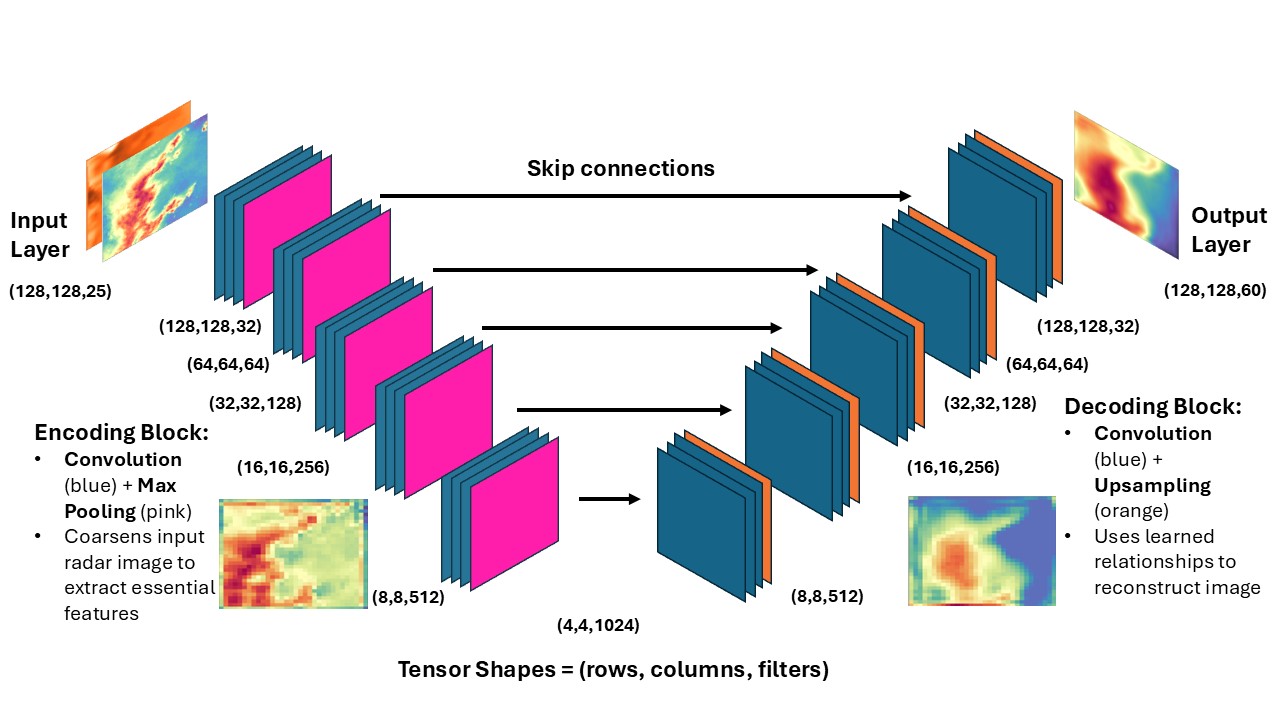}
    \caption{Model architecture diagram for the U-Net, with tensor shapes at each layer shown in parentheses. Feature maps from several convolutional/pooling layers are also shown throughout.}
    \label{fig:figure2}
\end{figure}

\subsection{Deep Learning Model}
We employ a U-Net \citep{RonnebergerUNet} to predict reflectivity evolution. U-Nets were originally developed for the task of image segmentation, which makes them suitable for taking input ``images", such as reflectivity maps, and producing new images as outputs. We use a specific variant of the U-Net, the U-Net 3+ \citep{HuangUNet3+}, which incorporates skip connections at all levels of the U-Net. Skip connections allow the output of certain layers in the model to be directly passed to the output of later layers, which helps to promote effective learning through more seamless backpropagation.

When training a U-Net, it is common to use loss functions such as mean squared error, which seeks to minimize the squared error between ground truth labels and predictions generated by the model. However, there are multiple shortcomings associated with using a loss function that outputs only a single deterministic value. Notably, for the atmospheric sciences, where probabilistic information is often crucial to guide decision making, training a model with a loss function that produces deterministic outputs (such as mean squared error) would not allow for inherent uncertainty quantification. To overcome this innate problem with conventional loss functions, we employ parametric regression, whereby the model learns to predict parameters of a probability distribution, rather than a single deterministic value \citep[e.g.,][]{BarnesSHASH, ChaseUVV}. Specifically, we use the SHASH distribution \citep{BarnesSHASH}, a distribution from the normal distribution family characterized by four parameters: location ($\mu$), scale ($\sigma$), skewness ($\gamma$), and tailweight ($\tau$). In the special case where the skewness parameter equals 0 and the tailweight parameter equals 1, the SHASH distribution converges to a normal distribution. From the parameters of the SHASH distribution, a unique probability distribution can be constructed, from which we can extract both deterministic (e.g., median) and probabilistic (e.g., 25th-75th percentile confidence intervals) information to generate reflectivity predictions. 

A negative log likelihood function is used as the loss function for the model:

\begin{equation}\label{eq:1}
    loss = -log_e(p)\\
\end{equation}

where \textit{p} is the probability of the ground truth value predicted by the SHASH distribution, characterized by the four parameters above. For a well-trained model, the negative log likelihood will be minimized (i.e., the median of the probability distribution will occur) near the ground truth value of reflectivity.

\subsection{Evaluation Metrics}
To evaluate the skill of our model predictions, we employ a variety of evaluation and explainability techniques. In this work, we present results on the probability integral transform D-statistic (PITD), the fractions skill score, and multiple explainability methods. Each is described in detail below.


The PITD statistic, and its associated PIT histogram, measure the probabilistic calibration of a model \citep{BarnesUQ,ChaseUVV}. An ideal model would generate predictions exactly equivalent to the ground-truth values; failing this, we would like to ensure that the model's predictions are well-calibrated (i.e. neither overconfident nor underconfident). 

The PIT histogram is computed by generating the probability integral transform of the predicted distributions and then determining where the ground truth values fall within the cumulative distribution function (CDF) of the predicted distributions.   A model with perfect probabilistic calibration will have well-dispersed predictions, meaning that each bin will have the same count of predictions (i.e. the ground truth values are evenly spread out throughout the predicted CDF) and the height of each bin will simply be $\frac{1}{B}$, where \textit{B} is the number of bins. Meanwhile, a poorly calibrated model might be overdispersed, which would result in a PIT histogram with a sharp peak in the middle (ground truth values always tend to fall in the center of the predicted distribution), or underdispersed, which would result in a U-shaped histogram (ground truth values tend to be on the edges of the predicted distribution). 

The PITD statistic is computed by first dividing the predicted distribution into some number of bins, \textit{B}. Thus, the PITD statistic measures the mean deviation of each bin's height from this perfect calibration ($\frac{1}{B}$) height. It is mathematically defined as below:

\begin{equation}\label{eq:2}
    PITD = \frac{1}{B}\sum_{k=1}^B({b_k-\frac{1}{B})^2}
\end{equation}

where \textit{$b_k$} is the frequency of occurrence in the \textit{k}-th bin. For a perfectly calibrated model, PITD would be 0. The PITD is analogous and statistically related to the rank histogram for assessing calibration of discrete probability distribution from ensembles of NWP models (\cite{Hamill2001}). We use the PIT histogram and PITD statistic to evaluate the median of the predicted SHASH distribution at each pixel, with evaluations stratified across lead times.

The fractions skill score (FSS) is a commonly used metric in the atmopsheric sciences for evaluating the skill of model predictions \citep{RobertsAndLeanFSS, ScheckFSS, NeckerFSS}. FSS computes the fraction of points on a grid where both the ground-truth and predicted values within a specified radius exceed a specified threshold. Because FSS is a neighborhood-based metric, it is useful to avoid the double penalty of slight displacements in space/time. For instance, if the forecast is accurate in magnitude but slightly offset in location, FSS will still score the forecast highly for skill at larger radii. FSS is formulated in terms of the Fractional Brier Score (FBS) and Worst-Possible Fractional Brier Score (WFBS) as follows:

\begin{equation}\label{eq:3}
    FSS = 1 - \frac{FBS}{WFBS}
\end{equation}
where $FBS = \frac{1}{I}\sum_{i=1}^I({NP_{i,f}}-NP_{i,o})^2$ and $WFBS = \frac{1}{I}\sum_{i=1}^I({NP^2_{i,f}}+NP^2_{i,o})$.

Here, ${NP_{i,f}}/{NP_{i,o}}$, respectively, are the neighborhood probability for a given radius of exceeding a threshold for the forecast and observation (prediction and ground-truth, in our case). An ideal FSS is 1, where both prediction and ground-truth exceed the threshold at the same frequency.
For more detail on FSS see \citet{NeckerFSS}.

\subsection{Explainability Techniques}

Permutation importance is one of the most commonly employed explainability methods in machine learning. Permutation importance involves taking the individual input variables, or features, and permuting their values to determine each one's impact on model performance. Here, we calculate permutation importance with respect to the SHASH loss function, so permutation importance is expressed as:
\begin{equation}\label{eq:4}
PI=loss(X_p)-loss(X)
\end{equation}
where $X_{p}$ is the permuted values of the original input variable $X$. If $loss(X_p) > loss(X)$, then the feature being examined improves the model's predictions, because performance degrades when its values are permuted. The overall permutation importance is calculated by taking the average change in loss over 100 iterations to ensure robust results.

To more deeply examine the impact of certain features on model predictions, we evaluate the accumulated local effects (ALE) of model predictions on each feature. ALE for a given feature is determined by selecting the feature of interest, and then binning that feature over its typical range. Once the bins have been assigned, for each bin we create two instances of the inputs for the ALE calculation: one instance where all pixels within that bin are assigned to the lower bound of the bin, and another instance where all pixels within that bin are assigned to the upper bound of the bin. Predictions are then generated for both input sets, and the difference between the input set with the upper bound of the bin and the input set with the lower bound of the bin is the \textit{local effects}. To obtain the \textit{accumulated local effects}, we sum the local effects over all bins and then center the effect by subtracting the mean local effect, thus providing a single curve that shows patterns of variation over the entire feature range. Mathematically, ALE is expressed as:

\begin{equation}\label{eq:5}
ALE(f) = \sum_{b=1}^B(\hat{y}_{upper} -\hat {y}_{lower}) - \frac{1}{N}\sum_{i=1}^n\sum_{b=1}^{B}(\hat{y}_{upper}-\hat{y}_{lower})
\end{equation}

where f is the feature for which ALE is being calculated, $\hat{y}_{upper}$ and $\hat{y}_{lower}$ are the predictions generated at the upper and lower bounds of each bin, respectively, and B is the total number of bins. After computing ALE, we sum over the full domain to obtain the accumulated effects over an entire image for one test set example.

We also create idealized input composites to explore subsets of the test set data where the model performs well or performs poorly. Input composites are created through the use of probability-matched means [PMMs; \citet{EbertPMM}]. PMMs are created from simple composite means of each feature within the test set; the first step after computing the simple composite mean is to "rank" all of the pixels within the simple composite according to their percentile within the composite. After the pixels are ranked accordingly, each one is re-assigned according to the corresponding percentile from images within the subset of the test set for which PMMs are being calculated (e.g. when computing PMMs for the best-performing cases, the 90th percentile of reflectivity within the simple composite mean image is re-assigned to whatever the 90th percentile at that pixel was for the subset of the best-performing cases). These PMMs are then used to explore the best-performing and worst-performing cases for the model within the test set.

An additional explainability method that we employ is novelty detection (\cite{McGovernBlackBox}), which aims to determine the most novel, or unexpected, images within the test set. To detect the most novel examples, we train a simple variational autoencoder (VAE; \cite{VAE}) on all examples within the test set. The VAE compresses its inputs into a lower-dimensional latent space and then reconstructs them with relationships learned during training. Unlike the U-Net, the VAE is trained with the inputs of the test set as both its feature and its target; the goal of the VAE is to most effectively reproduce the information contained within the test set inputs. After the VAE is trained, it is then tasked to predict the inputs of the test set. The reconstruction error is the error between the test set inputs and the predictions of the test set inputs generated by the VAE, and the cases for which the model has the highest reconstruction error are considered the most novel. 

Finally, we employ gradient-based methods (see \citet{McGovernBlackBox} for information on these method classes) to identify relevant spatial regions for the model in generating predictions. Gradient-based methods measure the strength of the backpropagation gradient through layers of the model, thus providing an estimate of sensitivity of model predictions to a particular region. A large gradient indicates that large amounts of information are being passed through the model at this point, suggesting a point (or region) that is important for the model's predictions. Similarly, Input*Gradient \citep{MamalakisXAI} helps to explain predictions from the model by multiplying the value of the gradient by the input at each point. The baseline gradient method provides an estimate of sensitivity to a particular input (i.e. how sensitive is the value of the output to the value of the input?), while the Input*Gradient method estimates feature attribution (i.e. how much contribution does the input feature make to the output at each point, relative to other features?). 

\section{Results}

\subsection{Case Study Creation}
To highlight the model's capabilities in predicting reflectivity evolution, we present results from a pair of case studies. The case studies presented below are on a 368x368 km domain, while the model generates predictions on 128x128 images. We use a larger domain for the case study results to capture a greater portion of the storm structure, particularly for storm morphologies with a larger spatial area such as a quasi-linear convective system (QLCS). To obtain predictions on a 368x368 grid, predictions are successively made on overlapping 128x128 images and then stepped over by 120 pixels. This step size is chosen to allow for a small, 8 pixel overlap between each individual image predicted by the model; in the regions where images overlap, the maximum value among the overlapped images is selected for the pixel output.

\begin{figure}
    \centering
    \includegraphics[width=0.8\linewidth]{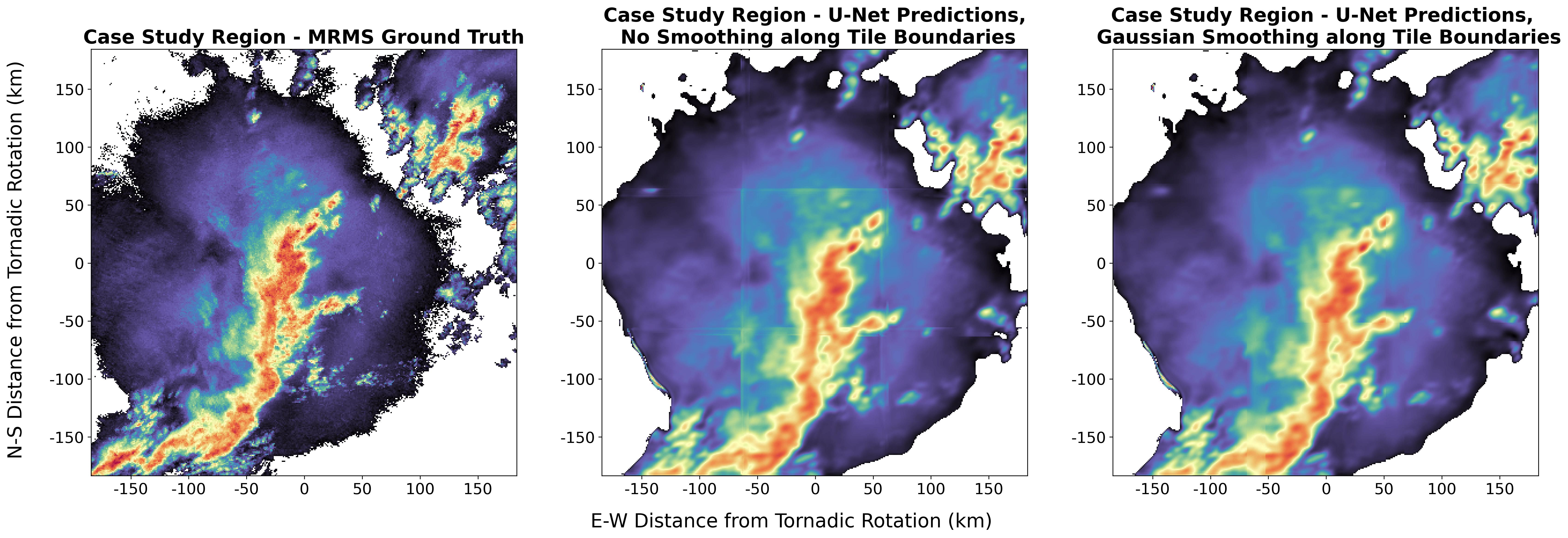}
    \caption{Example case study with MRMS ground truth (left) and the predictions generated by the U-Net (predictions shown are at the time of tornadogenesis, or a 2-minute prediction after the last input image). Both the MRMS ground truth and U-Net predictions are shown on a 368x368 grid, with the full U-Net predictions created by successively predicting on 128x128 image tiles, each of which overlaps by 8 pixels with the neighboring tiles. Predictions from the U-Net are shown with no smoothing along the overlapping tile boundaries (center) and with Gaussian smoothing along the overlapping tile boundaries (right). The Gaussian filter has a kernel standard deviation of 1.}
    \label{fig:figure3}
\end{figure}

Tiling images of the reflectivity fields together is a non-trivial task. Even with a small overlap between each image tile, discontinuities between adjacent images may occur. Figure~\ref{fig:figure3} shows ground truth MRMS data, along with image tiles of reflectivity predicted by the model. When tiling images together, seams between image tiles are evident when adjacent tiles are placed next to one another on the image with no filtering. To partially remedy these artifacts, we apply a Gaussian filter to the seams between images, which results in a smoother transition between adjacent tiles (See appendix A for discussion on additional tiling strategies that were tested in line with \citeauthor{HuangTiling} \citeyear{HuangTiling}.) The Gaussian filter applied to the image seams has a kernel standard deviation of 1 so as to avoid excessive blurring of the predictions along the tile boundaries. 

\subsection{Test Set Case Studies}

\begin{figure}
    \centering
    \includegraphics[width=0.8\linewidth]{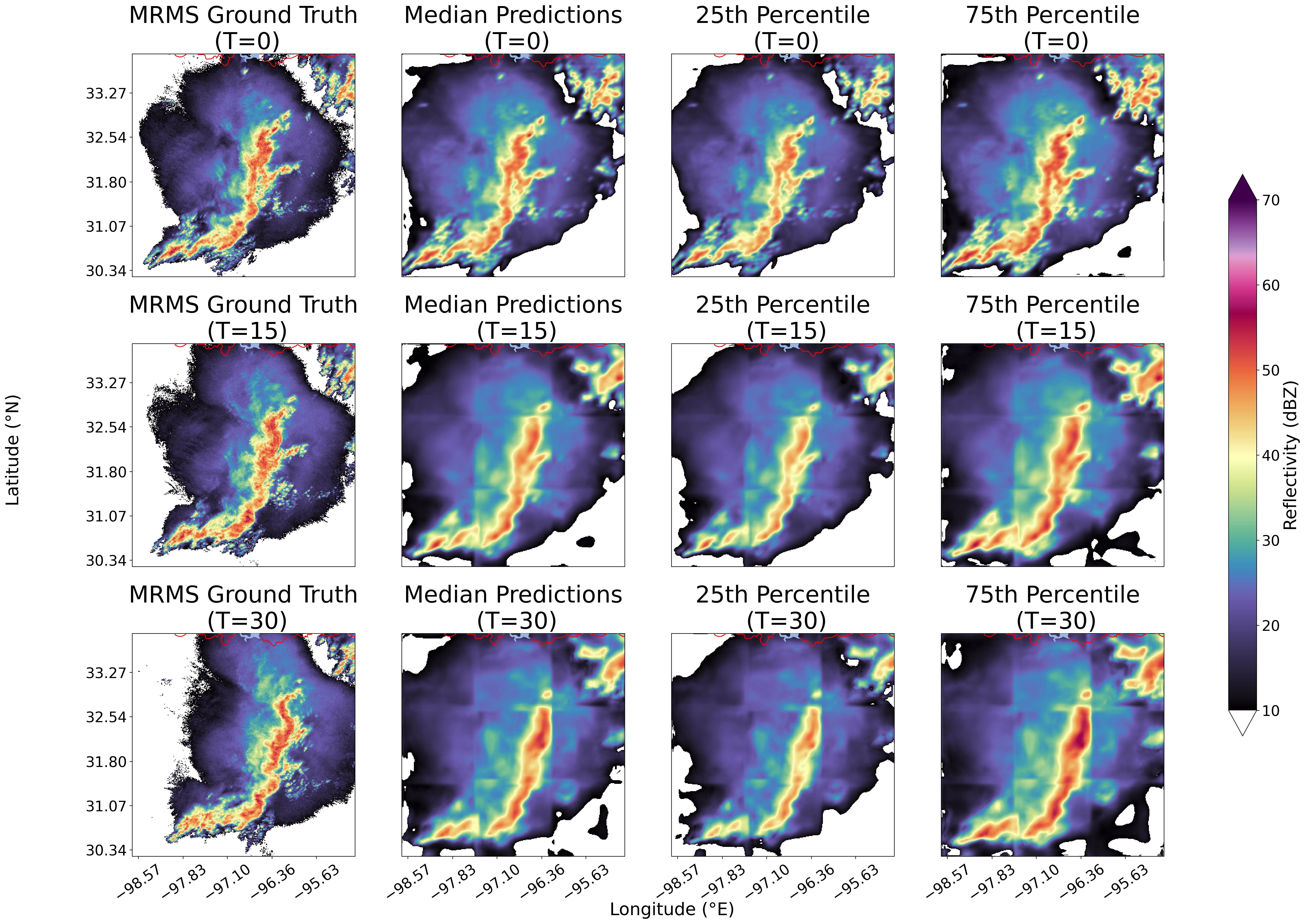}
    \caption{Case study results for a tornadic case at 0852 UTC 26 April 2024. From left to right: MRMS ground truth radar reflectivity, median predictions from the U-Net (i.e., median of predicted SHASH distribution at each point), 25th percentile predictions from the U-Net, and 75th percentile predictions from the U-Net. Ground truth and predictions shown at the approximate time of tornadogenesis (T=0; top row), 15 minutes following tornadogenesis (T=15; middle row), and 30 minutes following tornadogenesis (T=30; bottom row). Each tile has an overlap of 8 pixels on a side with its neighboring window. The red outlines denote state borders within the U.S.}
    \label{fig:figure4}
\end{figure}

\begin{figure}
    \centering
    \includegraphics[width=0.8\linewidth]{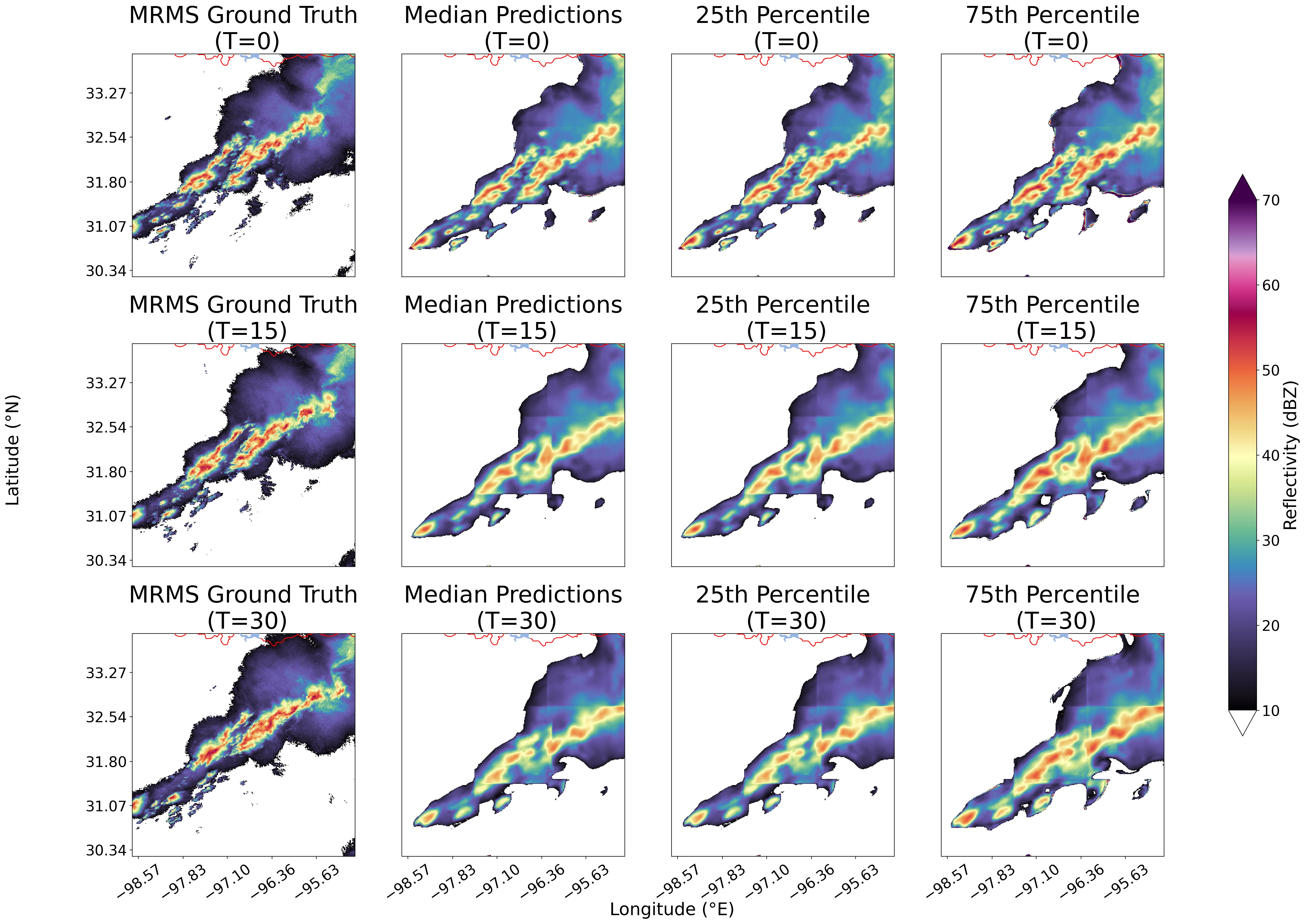}
    \caption{Same as Figure 4, but for a tornadic case at 1659 UTC 26 April 2024.}
    \label{fig:figure5}
\end{figure}

We begin our analyses with two case studies drawn from the testing set (Figures~\ref{fig:figure4} and ~\ref{fig:figure5}). The two test set cases analyzed occurred on 26 April 2024. A strong synoptic system produced favorable conditions for multiple rounds of severe weather throughout the Central United States. We selected this date for case studies due to the wide-ranging severe weather occurring throughout the U.S., and the variety of storm types that produced tornadoes, which provides a diverse set of cases for generating and evaluating model predictions.

The first case (Figure~\ref{fig:figure4}) involves a tornado that occurred in central Oklahoma, starting at 0852 UTC. This tornado originated from a QLCS with bowing segments and produced EF0 damage in Pottawatomie County, Oklahoma. Figure~\ref{fig:figure4} shows the real-time evolution of the tornado-producing storm on radar. At T=0, the predictions generated by the model match very well to the ground truth, with reflectivity intensity slightly higher in the main convective core at the 75th percentile. Over the next 30 minutes, the general structure of the QLCS is preserved, but the leading convection is not captured as well. By T=30, some of the finer-scale bowing segments of the QLCS are not as evident in the U-Net predictions, with less resolution evident near the semi-discrete cells at the northern and southern edges of the QLCS. However, the overall structure and intensity of the system is well-preserved even out to a 30-minute lead time.

Next, we examine another tornadic event from the same date (Fig.~\ref{fig:figure5}). This case involved a cluster of discrete cells in Hill County, TX. One of these discrete cells produced a brief EF2 tornado during its lifetime, before the storms began to grow upscale over the course of the next hour. As in Figure~\ref{fig:figure4}, the ground truth reflectivity from MRMS is presented alongside median, 25th percentile, and 75th percentile predictions from the U-Net's SHASH probability distribution output. The model captures multiple semi-discrete cells at the time of tornadogenesis and 15 minutes following tornadogenesis, which matches the presentation seen in MRMS. As the system evolves, the U-Net maintains the overall structure of the individual storms at each output percentile; the median predictions slightly underpredict reflectivity over time, while the 75th percentile predictions closely match the magnitude of reflectivity in the ground truth MRMS.

\subsection{Model Evaluation}

\begin{figure}
    \centering
    \includegraphics[width=\linewidth]{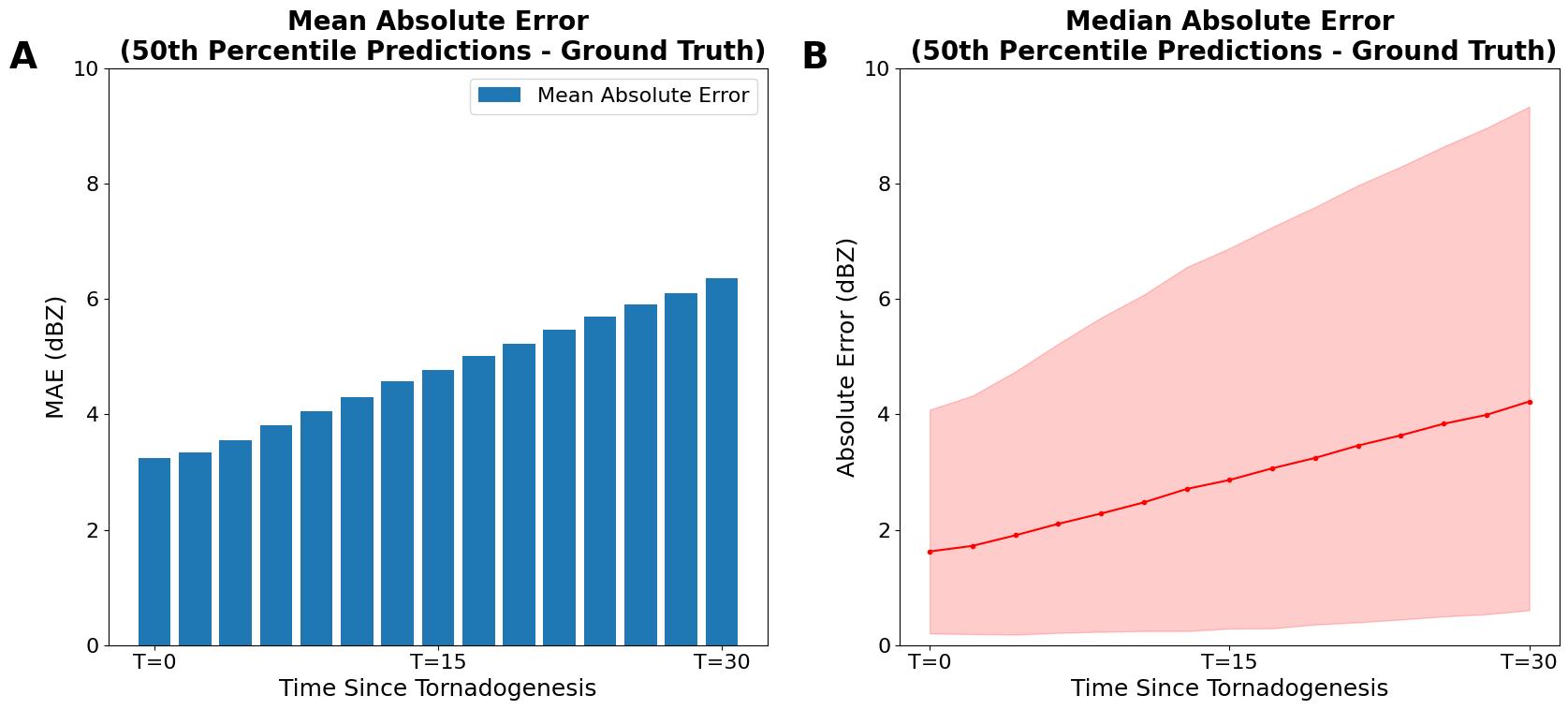}
    \caption{a) Test set mean absolute error between the 50th percentile predictions and the MRMS ground truth by timestep. b) Test set median absolute error between the 50th percentile predictions and the MRMS ground truth by timestep (line). Shading indicates the 25th and 75th percentiles of the error between the median predictions and the MRMS ground truth by timestep.}
    \label{fig:figure6}
\end{figure}

In addition to inspecting individual cases from our testing domains, we also perform evaluation of the model predictions in the aggregate. Figure~\ref{fig:figure6} demonstrates mean and median absolute errors between the median/50th percentile predictions for the test set, and the ground truth MRMS test set data. Figure~\ref{fig:figure6}a includes the mean absolute error (MAE) as a function of lead time, from tornadogenesis to 30 minutes following tornadogenesis. MAE gradually increases with lead time, with a mean error between predictions and labels of approximately 3 dBZ in the first predicted timestep, increasing to a mean error of just above 6 dBZ 30 minutes following tornadogenesis. 

Meanwhile, Figure~\ref{fig:figure6}b shows the median error between predictions and ground truth. The median error remains steadily lower than the mean error at all time steps. The 25th and 75th percentiles of the absolute error between median predictions and ground truth (25-75th percentile interval indicated by shading) demonstrate that errors are skewed towards the right tail of the distribution. This result indicates that there are  instances where the model has errors near or exceeding 10 dBZ (given that the 75th percentile of absolute error is just below 10 dBZ by T=30), which may be caused by the predictions blurring with lead time, leading to lower values of reflectivity in the predictions than are actually observed.

\begin{figure}
    \centering
    \includegraphics[width=\linewidth]{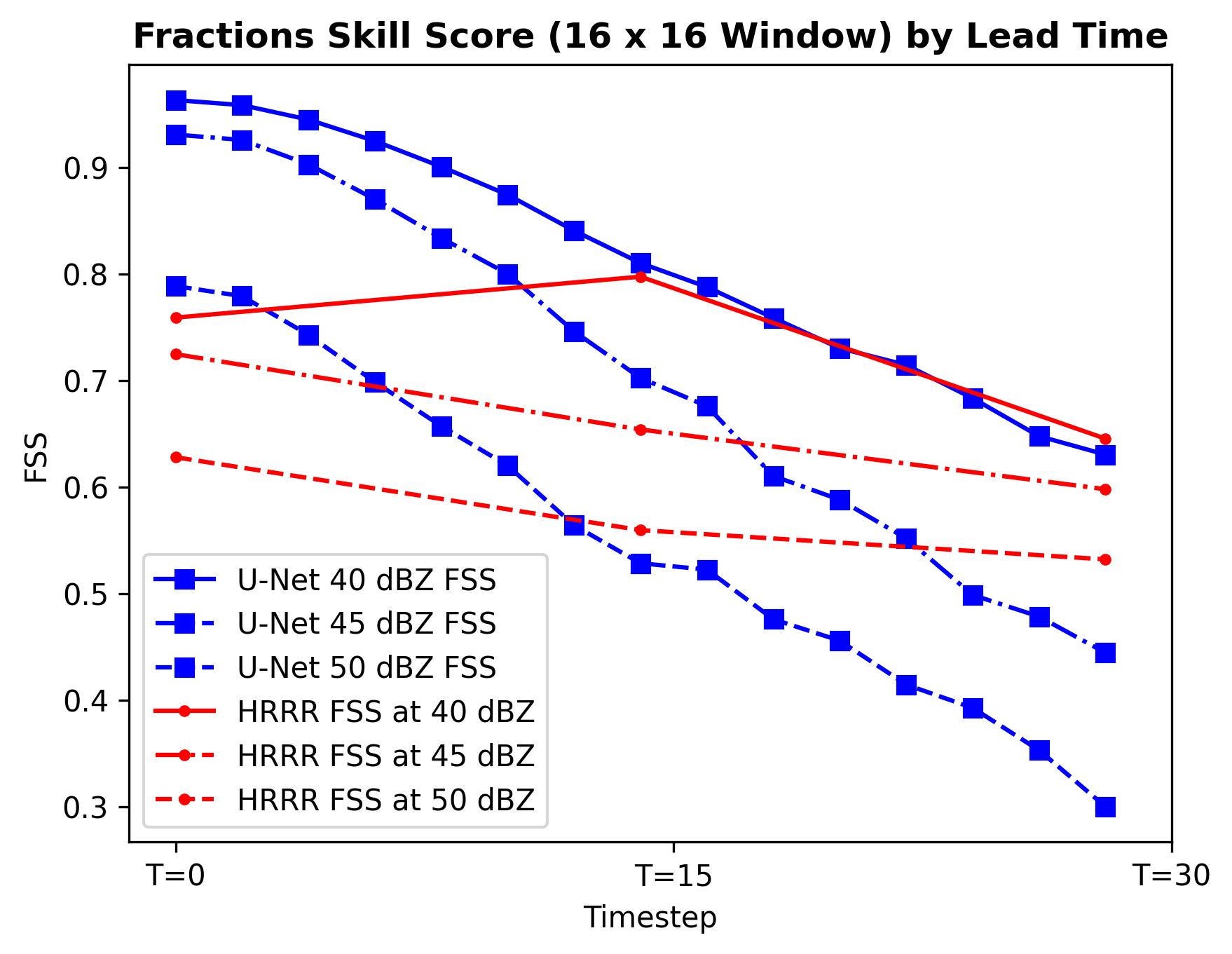}
    \caption{Fractions skill score (FSS) from the U-Net predictions (blue curves) and previous-hour HRRR forecasts (red curves) from the time of tornadogenesis (T=0) to 30 minutes following tornadogenesis (T=30). FSS is shown for the 40 dBZ, 45 dBZ, and 50 dBZ thresholds using a 16x16 neighborhood window. FSS for the U-Net is shown at 2-minute intervals based on the output timestep, while it is only plotted every 15 minutes for the HRRR because of the temporal resolution of sub-hourly HRRR data.}
    \label{fig:figure7}
\end{figure}

Beyond evaluating the absolute errors in the model's predictions, we also seek to determine whether our model is skillful relative to other forecasting tools. Figure~\ref{fig:figure7} presents a comparison of FSS for the U-Net and sub-hourly HRRR reflectivity forecasts. Sub-hourly HRRR forecasts are taken from the previous hour, and are compared to the U-Net predictions by identifying the closest storm in space from the previous hour's HRRR forecast to the tornado-producing storm in MRMS. Because the sub-hourly HRRR forecasts are only available at 15-minute intervals, we match storms between MRMS and the HRRR and round to the nearest 15-minute interval to evaluate HRRR forecasts at approximately 0, 15, and 30 minutes following tornadogenesis (by doing so, the maximum time offset between the sub-hourly HRRR forecast and the intended 15-minute comparison interval is 7 minutes). If no such storm existed in the previous-hour HRRR forecast run, the case is discarded to avoid unfairly penalizing the HRRR forecast in the comparison. 

The FSS for the median U-Net predictions and for the HRRR within a 16x16 grid point window as a function of lead time are shown in Figure~\ref{fig:figure7}. FSS was also calculated at smaller (4x4, 8x8) and larger (32x32, 64x64, and 128x128) window sizes, with similar qualitative performance across window sizes. At the time of tornadogenesis, the U-Net demonstrates strong skill at all thresholds from 40-50 dBZ, outperforming the HRRR at each of these thresholds. U-Net skill decreases somewhat with lead time, particularly at the higher reflectivity thresholds. Meanwhile, skill from the HRRR decays more slowly, and tends to be slightly higher than the U-Net by 20 minutes following tornadogenesis. By T=30, the sub-hourly HRRR forecast and the U-Net perform comparably at the 40 dBZ threshold, while the HRRR performs better at higher reflectivity thresholds.

\begin{figure}
    \centering
    \includegraphics[width=\linewidth]{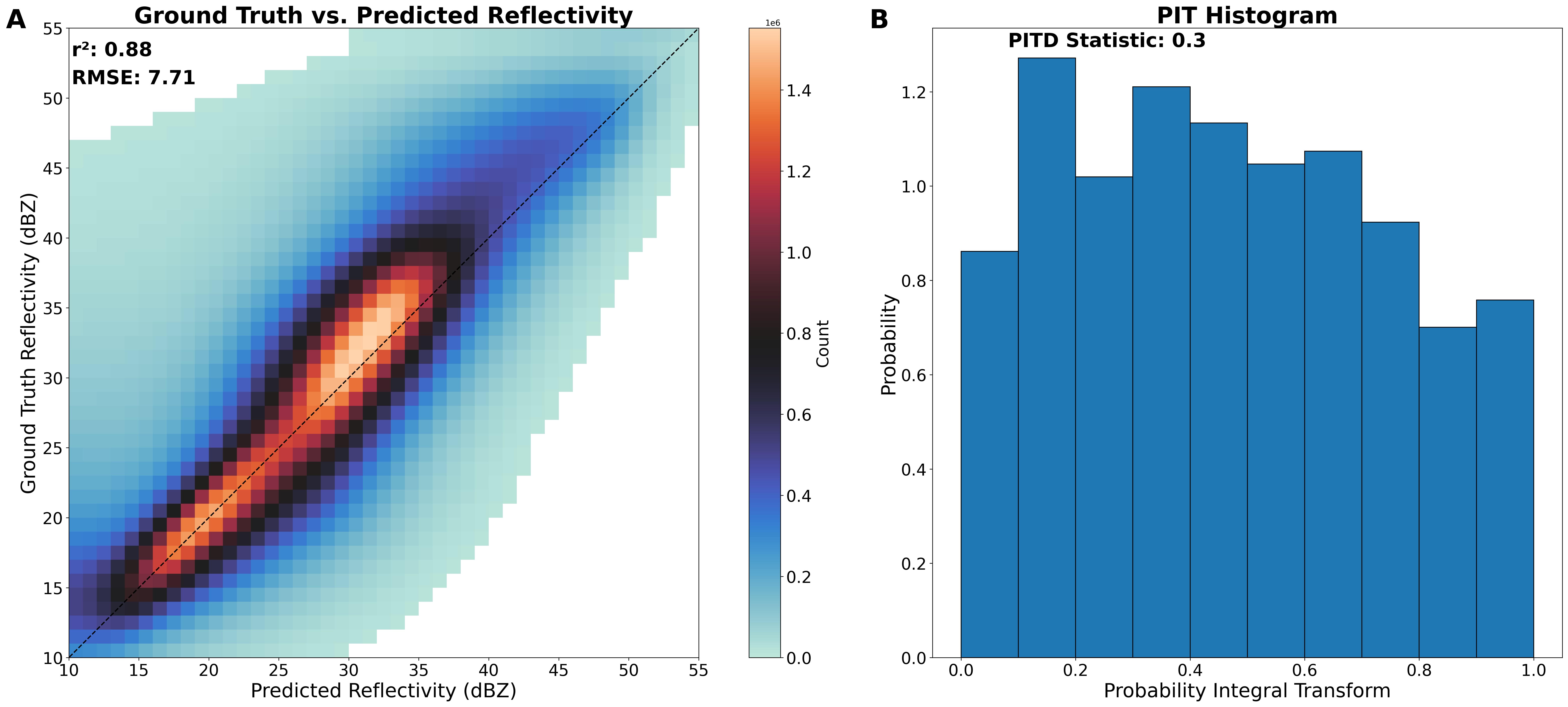}
    \caption{a) Pixel-wise comparison of the U-Net predicted reflectivity with the MRMS ground truth reflectivity. Pearson correlation coefficient and root-mean-squared-error included at top left. b)  PIT histogram to measure probabilistic calibration of model. PITD statistic included at top left.}
    \label{fig:figure8}
\end{figure}

Lastly, we examine the correspondence between the ground truth reflectivity data and the U-Net predictions in the aggregate. Figure~\ref{fig:figure8}a shows the pixelwise correspondence between ground truth reflectivity and predicted reflectivity in the test set. The correlation between ground truth and predicted reflectivity pixels is strong, with an $r^2$ value of 0.88 for the test set. The RMSE of 7.71 dBZ, when evaluated along with the MAE results shown in Figure 6, indicates some large-magnitude pixelwise errors between the ground truth and predictions, but still indicates good performance by the model in the aggregate. The distribution of predicted vs. ground truth reflectivity demonstrates that the model-predicted reflectivity generally corresponds quite well with the ground truth reflectivity at lower dBZ values. At dBZ values above 30, reflectivity is often slightly underpredicted by the model.

Figure~\ref{fig:figure8}b shows the PIT histogram and PITD statistic for the U-Net to evaluate the probabilistic calibration of its predictions. For a perfectly calibrated model, each bin would have a height of 1, while a model with poor calibration would have bin heights well removed from 1. The PIT histogram indicates reasonable probabilistic calibration of the model - most bin heights are slightly removed from the perfect calibration height of 1. Additionally, the higher bin heights tend to be concentrated in the middle of the PIT histogram, rather than on the edges, indicating that the model is overdispersive, with ground truth values typically falling near the middle of predicted distributions.

\subsection{Model Explainability}

\begin{figure}
    \centering
    \includegraphics[width=\linewidth]{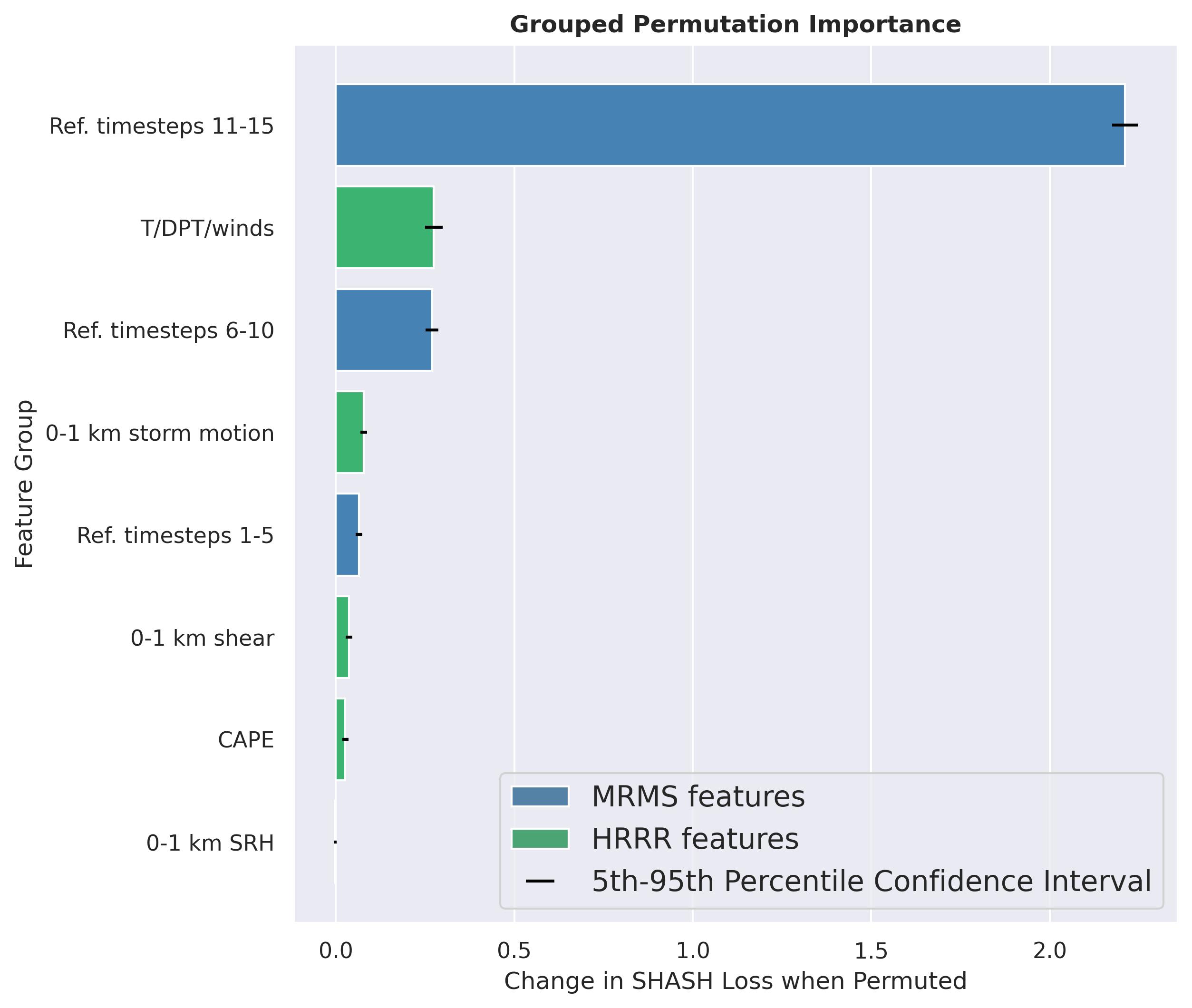}
    \caption{Grouped permutation importance for all features used for training the model. Bars are colored based on whether the feature group consists of features from MRMS or from the HRRR. The 5th-95th percentile confidence interval of the grouped permutation importance is included in black on each bar. Results were calculated by averaging the results of 100 iterations over the test set.}
    \label{fig:figure9}
\end{figure}

In addition to evaluating the quality of the model predictions, we also aim to provide thorough explanations of why the model produces predictions as it does, and to identify features and/or regions that contribute significantly to predictions. Figure~\ref{fig:figure9} shows the grouped permutation importance for the U-Net, with features grouped into reflectivity time steps at different times relative to tornadogenesis and multiple different categories of HRRR variables (including thermodynamic and kinematic variables, near-surface variables, etc). Permutation importance was calculated by averaging the results of 100 iterations over the test set. Mean differences in the importance values calculated were on the order of $10^{-2}$ or less across iterations, indicating robustness across trials.

The most important feature group in influencing model predictions is the reflectivity time steps closest to tornadogenesis (time steps 11-15, or 2-10 minutes prior to tornadogenesis). This group increases the loss value when permuted approximately 7x more than the next most important feature groups, the HRRR-forecast temperature/dew point/winds and the reflectivity time steps from 10-20 minutes prior to tornadogenesis. Earlier reflectivity time steps and other variables from the HRRR had a very small influence on the model's predictions on average.

\begin{figure}
    \centering
    \includegraphics[width=\linewidth]{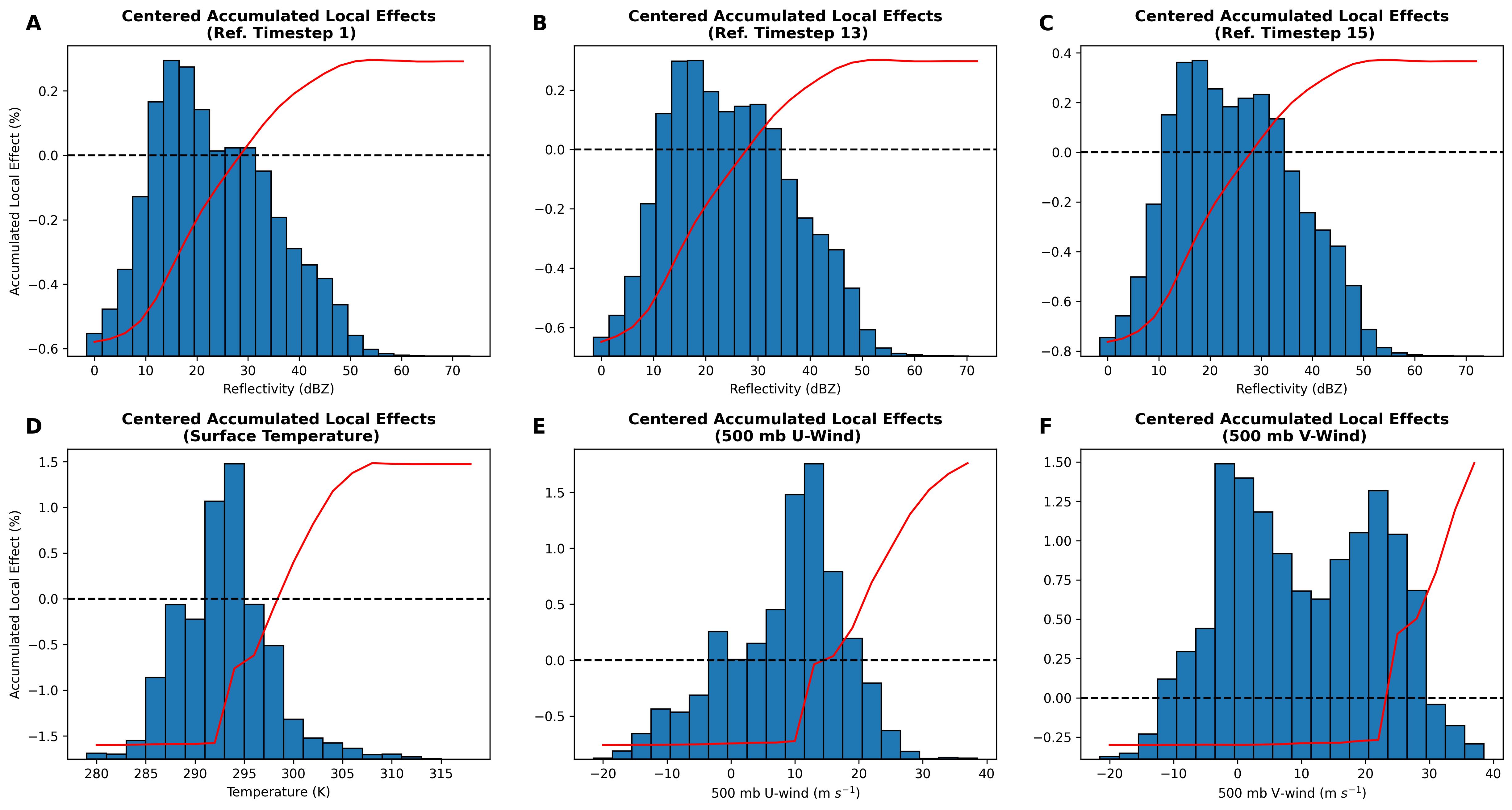}
    \caption{ALE plots for features from the training dataset: a) the MRMS reflectivity image 30 minutes prior to tornadogenesis, b) the reflectivity image 6 minutes prior to tornadogenesis, c) the MRMS reflectivity image 2 minutes prior to tornadogenesis, d) the previous-hour HRRR-forecast 925 mb temperature, e) the previous-hour HRRR-forecast 500 mb U-wind, and f) the previous-hour HRRR-forecast 500 mb V-wind. Line plot shows the ALE by feature bin. Histogram shows the distribution of the variable's values within the test set. The y-axis plots the ALE values (i.e. in each bin of the feature, how much do the predictions locally change?).}
    \label{fig:figure10}
\end{figure}

Figure~\ref{fig:figure10} compares the ALE of six features from the dataset, three of which are MRMS reflectivity time steps and three of which are HRRR forecast variables. The reflectivity time steps examined correspond to 30 minutes (Figure 10a), 6 minutes (Figure 10b) and 2 minutes (Figure 10c) prior to tornadogenesis. Each of the reflectivity time steps produced similar accumulated local relationships with the output predictions; for lower values of reflectivity in the reflectivity feature, the accumulated local effect is negative, meaning that predictions in this region are lower than the average baseline. Conversely, as the feature values increase, the predicted values also increase relative to the average baseline.

For the HRRR features, we see similarly monotonic relationships in the accumulated local effects. However, the increase in the local effects is more dramatic as the input feature values increase, particularly for the near-surface temperature and the 500 mb V-wind. These results suggest that, for most of the range of feature values, temperature and V-wind do not substantially contribute to the output predictions, as the local effects are small. But, for extreme values of the predictors, the corresponding effects on the predictions are much larger in comparison, suggesting important relationships in particular subsets of the feature space.

\begin{figure}
    \centering
    \includegraphics[width=\linewidth]{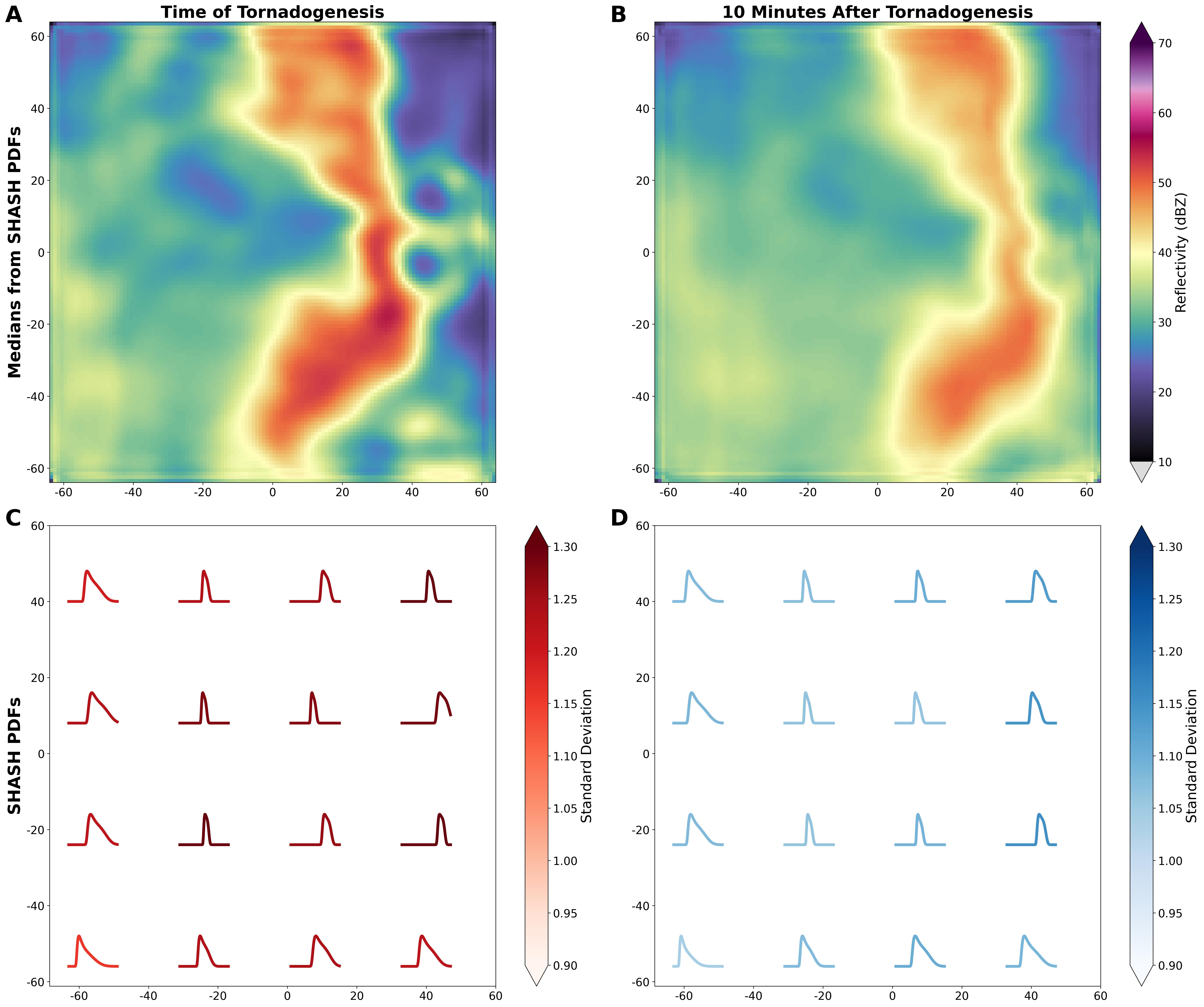}
    \caption{Median predictions from the SHASH probability distribution (top) and the probability distribution functions (PDFs; bottom) from which the median predictions are drawn. Predictions and PDFs shown at the time of tornadogenesis (left; subpanels A and C, respectively) and 10 minutes following tornadogenesis (right; subpanels B and D, respectively) for a select case. The PDFs are colored according to the value of the standard deviation parameter from the SHASH distribution at each point shown.}
    \label{fig:figure11}
\end{figure}

We also examine the spatial patterns of probability distribution functions (PDFs) generated by the model. Figure~\ref{fig:figure11} shows predictions generated by the model for a tornado produced by a linear system, both at the time of tornadogenesis and 10 minutes following, along with the accompanying field of PDFs at each timestep (sub-sampled for visual clarity). At the time of tornadogenesis, the model produces probability distributions suggesting higher reflectivity (i.e. higher predicted mean values, with peak of distribution shifted further to the right) in the convective areas of the prediction domain. The model also assigns higher values of standard deviation to the convective areas of the storm, indicating greater uncertainty in these regions. By 10 minutes following tornadogenesis, the tornado-producing storm in the predictions has moved off to the east. At this time, the highest mean values in the PDFs are still associated with the convective areas of the storm along the eastern edge of the prediction domain. However, the standard deviation values predicted by the model have decreased somewhat, indicating that the model is not as confident in predicting uncertainty in the distribution as predictability is lost further in time from the inputs it received.

\begin{figure}
    \centering
    \includegraphics[width=\linewidth]{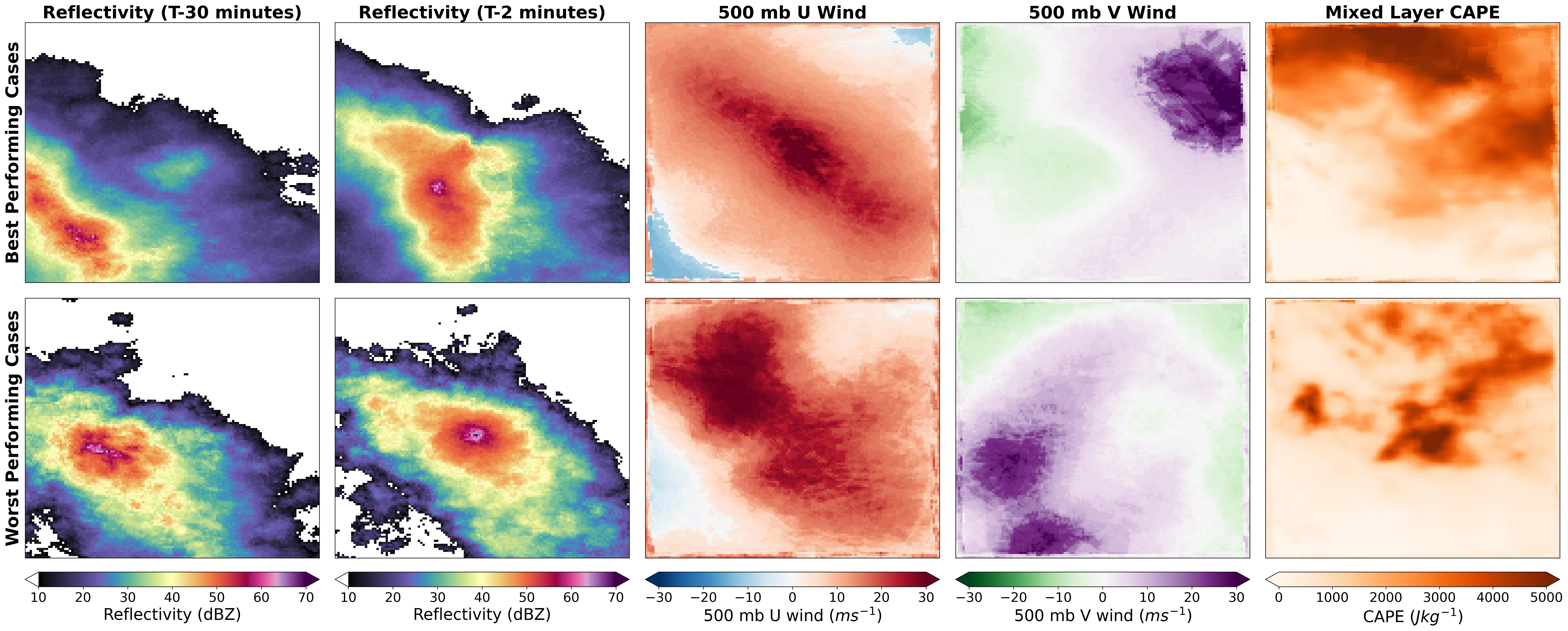}
    \caption{Composites of input features for the 25 test set cases with the lowest RMSE (i.e. best-performing cases; top) and the 25 test set cases with the highest RMSE (i.e. worst-performing cases; bottom). Composites are presented for (from left to right): reflectivity 30 minutes prior to tornadogenesis, reflectivity 2 minutes prior to tornadogenesis, 500 mb U-wind, 500 mb V-wind, and CAPE.}
    \label{fig:figure12}
\end{figure}

Beyond considering the impact of features on individual predictions, we also explore the aggregate effect of different features on the model's predictions. Figure~\ref{fig:figure12} showcases the PMM composites of different features for the 25 test set cases with the lowest RMSE (i.e. the 25 best-performing cases for the model) and the 25 test set cases with the highest RMSE (i.e. the 25 worst-performing cases for the model). To focus strictly on the impacts of storm-scale characteristics on the quality of the model predictions, rather than on the impact of shifting the storms within the domain, we compute the RMSE difference only for those samples which are centered on the storm, rather than including all of the randomly shifted samples as well; composites are then generated from this subset of storm-centered images. 

For the reflectivity feature from 2 minutes prior to tornadogenesis, the composite for the low RMSE (i.e. best-performing) cases shows a relatively smooth reflectivity field over a large region, suggesting that these low RMSE cases tend to be associated with tornadic events from storm morphologies with more large-scale organization, such as QLCSs. Meanwhile, the high RMSE (i.e. poorly-performing) composite shows a reflectivity field much more characteristic of a discrete supercell, which would suggest that the model struggles more with capturing smaller-scale evolution. The composite features for reflectivity from 30 minutes prior to tornadogenesis indicate similar results; the low RMSE composite for this feature shows a larger, smoother reflectivity field, while the high RMSE composite shows a weaker discrete morphology 20 minutes prior to tornadogenesis, again suggesting that the model is better able to capture trends associated with larger-scale organization. 

For the HRRR feature composites, both the low and high RMSE cases for the 500 mb U-wind indicate primarily westerly winds with fairly similar structures. The 500 mb V-wind composites show a dipole structure of northerly and southerly winds for the low RMSE cases, and a more consistent pattern of only southerly winds for the high RMSE cases. Finally, the composites of CAPE from the low vs. high RMSE cases indicate a roughly meridional buoyancy gradient, with CAPE generally increasing from south to north, in the low RMSE cases. For the higher RMSE cases, the spatial pattern of CAPE forecast by the HRRR is more diffuse, with a much noisier meridional gradient evident. 

\begin{figure}
    \centering
    \includegraphics[width=\linewidth]{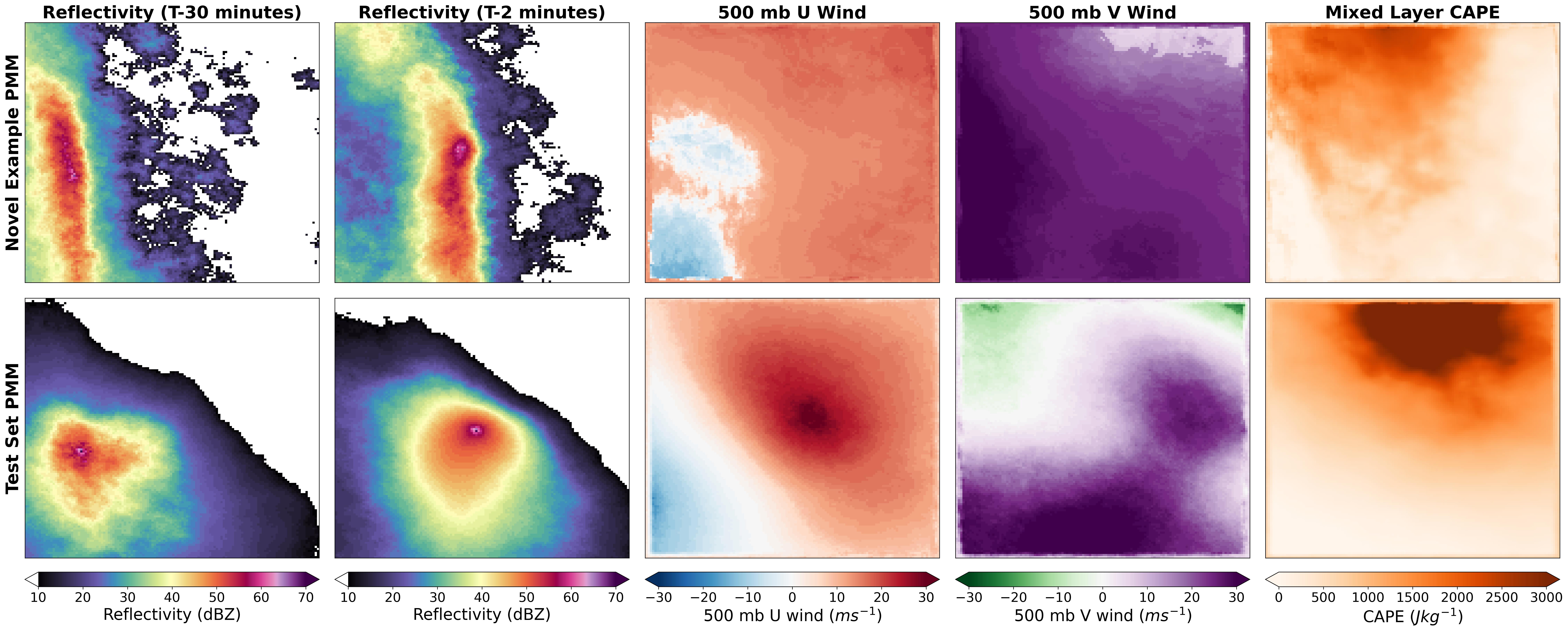}
    \caption{Same as Figure 12, but for the 25 most novel cases in the test set (top) and for the test set as a whole (bottom).}
    \label{fig:figure13}
\end{figure}

We use a similar method to explore the most novel cases in our test set (Figure~\ref{fig:figure13}). After training a VAE on the test set, the most novel cases are determined by identifying those with the highest reconstruction error when the trained VAE makes predictions on the test set. PMMs of the most novel examples, along with PMMs for the overall test set, are shown in Figure~\ref{fig:figure13}. For the reflectivity features, the most novel examples feature a highly linear tornado-producing storm, suggesting that highly-organized QLCSs may account for some of the most novel examples in the test set. Meanwhile, the novel examples of U-wind indicate a relatively weak field of westerlies everywhere in the vicinity of the tornado-producing storm, compared with the dipole structure present in the overall test set composite. The novel V-wind examples contain a strong field of southerly winds throughout, which is again contrasted with a more dipole-like structure in the overall test set PMM. Finally, both the novel examples and the overall test set composite indicate a roughly meridional gradient of CAPE, similar to the low RMSE composites presented in Figure~\ref{fig:figure12}. 

\begin{figure}
    \centering
    \includegraphics[width=\linewidth]{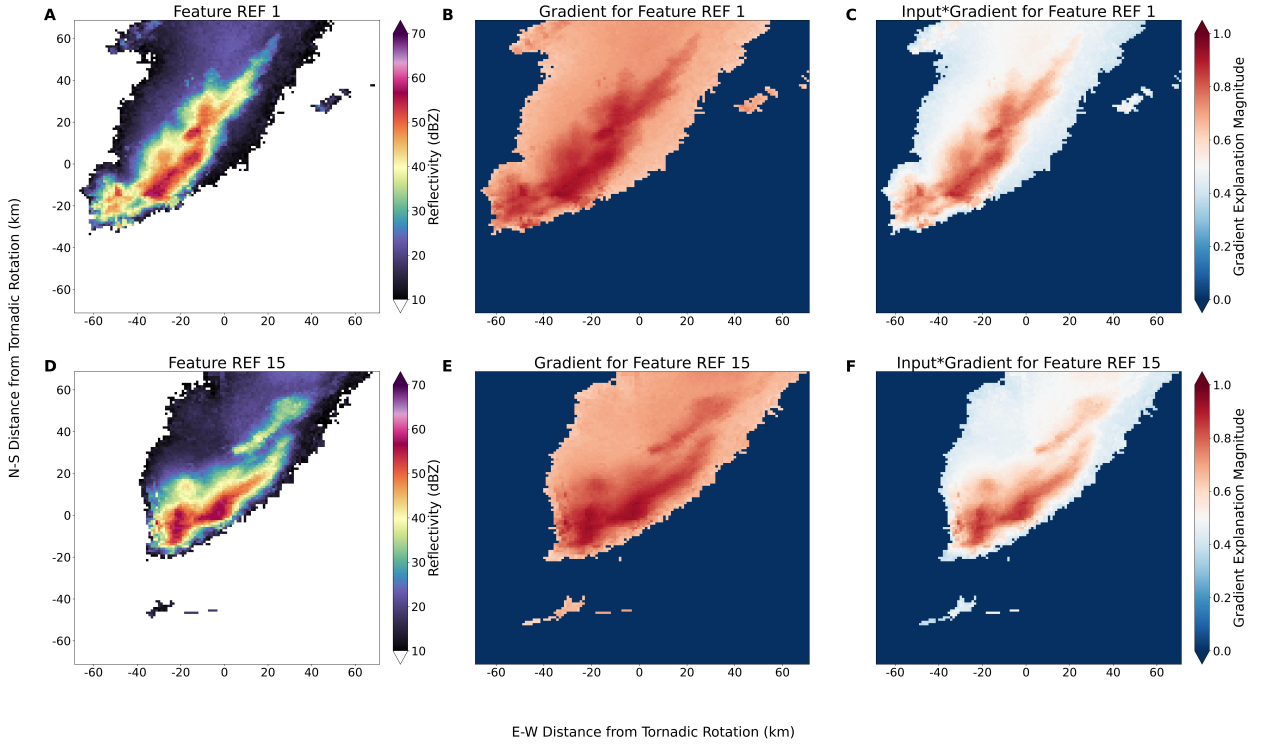}
    \caption{a) The first input timestep (i.e. 30 minutes prior to tornadogenesis) of reflectivity imagery for a case on April 26th, 2024 at 2018 UTC. b) The gradient explanation, measured by the strength of the backpropagation gradient at each point, for the reflectivity image. c) The input*gradient explanation (calculated by multiplying the value of the input times the value of the backpropagation gradient at each point) for the reflectivity image. The same results are presented in d/e/f, but for the final input timestep prior to tornadogenesis (i.e. 2 minutes prior).}
    \label{fig:figure14}
\end{figure}

Lastly, we employ gradient-based methods to visually examine relevant regions influencing model predictions. Both the vanilla gradient (indicating local sensitivity to feature values) and the input*gradient (indicating the degree of feature attribution) are presented. Figure~\ref{fig:figure14} shows the gradient explanation and the input*gradient explanation for the first reflectivity input timestep (i.e., 30 minutes prior to tornadogenesis) and the final reflectivity timestep (2 minutes prior to tornadogenesis) of a tornadic discrete cell case on April 26th, along with the reflectivity values accompanying the explanation. The magnitude of the backpropagation is clearly strongest in the regions of the convective core of the storm, indicating that the most information is being passed through the model from this region. Meanwhile, very low values of the gradient correspond to regions of low reflectivity, indicating that the model is not locally sensitive to changes outside of the convective storm. For the input*gradient, this relationship is amplified - the highest values of input*gradient correspond to the convective core of the storm, indicating that this region provides the most contribution to the output of the model when making predictions for this case.

2 minutes prior to tornadogenesis, the backpropagation gradient remains strongest in the regions of highest reflectivity, indicating that the model takes the most information from these regions (Fig.~\ref{fig:figure14}e). Additionally, weak convection that existed to the east of the main cell in the first timestep has dissipated, and a new region of weak convection has appeared to the south of the parent cell. The model also has a relatively strong backpropagation gradient in this region, suggesting that the presence of new convection, as isolated and weak as it appears, could influence the model's predictions following tornadogenesis. Similarly as the first reflectivity timestep, input*gradient shows the strongest explanation in the convective regions of the storm, with weaker feature attributions coming from other regions. Like the vanilla gradient, it also identifies the weak convective echoes to the south of the parent storm as a region of potential explanation.

\section{Discussion and Conclusions}

We introduce an explainable, probabilistic deep learning system for radar nowcasting in tornadic thunderstorms. While other deep learning nowcasting systems exist for predicting precipitation/radar reflectivity, this work provides a probabilistic system with a rigorous explainability routine. We examined multiple case studies from a central U.S. tornado outbreak, finding that the deep learning model is capable of producing realistic predictions of reflectivity evolution out to 30 minutes in the future, despite blurring of the forecast at longer lead times. We also demonstrated that our model provides comparable skill to HRRR reflectivity forecasts. Finally, we presented several explainability metrics to further understanding of important variables and regions in performing radar nowcasting with this probabilistic system.

Despite the progress presented here, there are substantial potential improvements that can be made to the model in the future. Firstly, additional changes could be made to the model architecture to promote increased sharpness in the model predictions (see \citeauthor{EbertUphoffSharpness} \citeyear{EbertUphoffSharpness} for discussion of sharpness in AI applications for meteorology). One weakness of the model predictions currently is a blurring of predictions, whether that be along boundaries when storms are clustered together, or as lead time increases. An attempt to improve sharpness could involve further modifications to the loss function used to train the model, e.g., by adding a term to the loss function that penalizes the model for blurry predictions. Additionally, when creating case studies by combining multiple image tiles together, there are often discontinuities between the predictions in adjacent image tiles, resulting in linear artifacts along the tile borders. While these discontinuities are improved by applying techniques such as Gaussian smoothing along the tile borders, the artifacts do remain to a degree, and warrant further exploration for improvement. 

We also sought to apply the model in transfer domains to evaluate its generalizability. We tested the model's capabilities on nowcasting radar reflectivity in tropical cyclones which produced tornadoes. Tornadoes originating from tropical cyclones (commonly TCTs) are often accompanied by radar signatures distinct from those found in overland tornadoes (see \citeauthor{SchneiderSharpTCT} \citeyear{SchneiderSharpTCT} and \citeauthor{SPCTCTs} \citeyear{SPCTCTs} for discussion on the differences in radar presentation between TCTs and overland tornadoes). Therefore, because of their distinct characteristics, nowcasting radar associated with TCTs would be a potentially useful transfer domain. However, we found generally poor performance from our model when applied to a separate TCT test set, with very blurry predictions and low skill (not shown). Further work could be done in the future to improve model performance in this or other transfer domains.

Additionally, the probabilistic calibration of the model indicates some potential for future improvement. The current PIT histogram shows a tendency towards slightly underconfident predictions, with outputs that are too spread out compared to the true degree of spread found in the ground truth data. Post-training probabilistic calibration (for instance, through fine-tuning of the individual SHASH distribution parameters) could be applied to improve the robustness of the model in generating distributions that best capture the degree of spread of the ground truth reflectivity fields.

Beyond using composite reflectivity as a predictor, other radar-based features such as differential reflectivity (ZDR) and specific differential phase (KDP) could be incorporated into the model. Both of these features are used to interpret tornado potential, distinguishing between tornadic and non-tornadic storms \citep{Brune2025Polarimetric}. Including these features in the model could enhance the model's capability to predict the evolution of storm structure.

Finally, there are opportunities to incorporate new model types entirely for radar nowcasting. Notably, diffusion models have shown success in meteorological applications recently \citep{Chase2025Diffusion,Asperti2025Precipitation}. While a U-Net is designed to segment an image into its constituent parts, diffusion is explicitly applied for image generation, which could lead such a model to be more successful in generating high-resolution images of radar reflectivity. Additionally, the high importance of later reflectivity time steps, compared to the much lower importance for earlier time steps, indicates that the model has not fully learned the temporal relationships across time steps. To improve the learning of information across time, attention-based architectures could be incorporated which learn to focus on particular regions across multiple time steps. Exploring additional model architectures might also be beneficial for making the model more generalizable, which could improve performance when nowcasting reflectivity in TCTs or in other domains.

This work represents a promising first step towards an operational deep learning model to nowcast radar evolution in tornadic storms. While potential improvements to the model architecture or feature set used still exist, the results presented here indicate the capability of deep learning models to generate skillful predictions of radar evolution in a nowcasting setting. Additional performance gains could move the model towards operational use in the future.

\clearpage

\acknowledgments
This material is based on work supported by the U.S. National Science Foundation under Grant No. RISE-2019758 and IIS 2331908. The lead author thanks Dr. Randy Chase for insightful explanations of the SHASH implementation, and Dr. Marina Vicens Miquel for assistance in acquiring MRMS data used in the work. Computing for this project, including data pre-processing, model training, and analysis, was performed at the OU Supercomputing Center for Education \& Research (OSCER) at the University of Oklahoma (OU).

\datastatement
Both the MRMS and the HRRR dataset are publicly available on AWS buckets. The MRMS bucket can be accessed at this link: https://registry.opendata.aws/noaa-mrms-pds/. The HRRR bucket can be accessed at this link: https://registry.opendata.aws/noaa-hrrr-pds/. Scripts used for preprocessing, model training, and supporting scripts are available at https://github.com/erickson-wx/TorU. 

\appendix
\appendixtitle{Tiling Strategies}

\renewcommand{\thefigure}{A\arabic{figure}}
\setcounter{figure}{0}
As discussed in section 4a, effectively tiling multiple adjacent reflectivity images into a larger domain is a non-trivial task. Over the course of this work, we tested numerous strategies to merge adjacent tiles into a seamless case study image. 

\begin{figure}
    \centering
    \includegraphics[width=\linewidth]
    {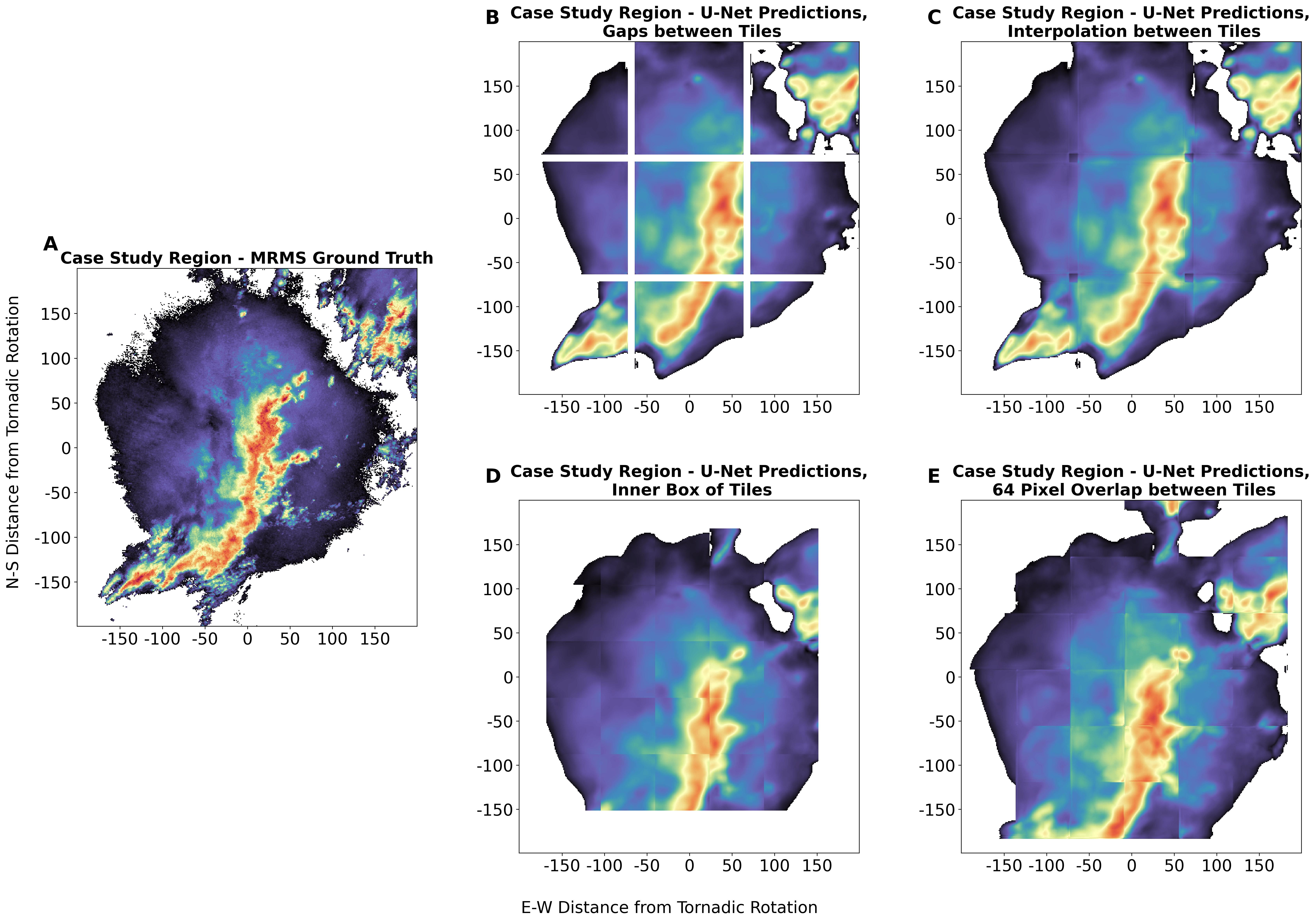}
    \caption{An overview of the different image tiling strategies tested for the case study creation. a) The ground truth MRMS reflectivity for one case drawn from the test set. b) U-Net predictions on a 400x400 km grid, with 8 pixel gaps separating each adjacent 128x128 image tile. c) The same predictions on a 400x400 km grid, with liner interpolation between adjacent tiles. d) U-Net predictions on the 400x400 km grid, with only the inner 64x64 region of pixels preserved for each image tile. e) U-Net predictions on the 400x400 km grid, with an overlapping window size of 64 pixels between adjacent tiles (the maximum value at each pixel is taken where tiles overlap).}
    \label{fig:fig1_appendix}
\end{figure}

In addition to the strategies presented in 4a (an 8-pixel overlap between adjacent tiles, with and without Gaussian smoothing), we also present a variety of alternative strategies in Figure \ref{fig:fig1_appendix}. Additional strategies implemented involved modifications to the pattern of adjacent tiles, such as leaving a small gap between adjacent tiles and interpolating linearly between them (shown in Figure A1b/c; functionally the opposite of including a small overlap between tiles) or using more advanced distance weighting methods such as Tukey/Hann blending (results not shown). We also tested multiple different degrees of overlap between tiles. One such strategy involved generating predictions on the full 128x128 image for each tile and only retaining the inner 64x64 grid points of the image (Figure A1d), or using a 64-pixel overlap between each adjacent tile and taking the maximum value at each pixel (Figure A1e). After testing a variety of different strategies for case study creation, we decided to use the method presented in section 4. 

\bibliographystyle{ametsocV6}
\bibliography{references}

\end{document}